\documentclass[12pt]{article}
\usepackage{amssymb}
\usepackage{epsfig}

\advance\textheight by \topskip
\input{tcilatex}

\begin{document}

\title{\textbf{Nucleosynthesis Without a Computer}}
\author{V. Mukhanov \\
LMU, Sektion Physik, Theresienstr.37, \ 80333 Muenchen}
\maketitle

\begin{abstract}
I derive completely analytically the time evolution and final abundances of
the light elements (up to $^{7}Be$) formed in the big-bang
nucleosynthesis.This highlights an interesting physics taking place during
the formation of light elements in the early universe.
\end{abstract}

\section{Introduction}

The most occurrent chemical element in the universe is hydrogen. It
constitutes nearly three quarter of all baryonic matter. The next mostly
wide spread element \textit{Helium-4}$,$ constitutes about 25\%. The other
light elements and the metals occur very rare. Very simple arguments lead to
the conclusion that it is very unlikely that $^{4}He,$ deuterium $\left(
D\right) $ and other light elements could be burned in the stars (see, for
instance, \cite{W},\cite{KT}). Therefore the only sensible explanation of
their abundance is that they were produced in the very early universe. It is
clear that the essential amount of the helium could not be formed before the
temperature dropped below its binding energy $\sim 28$ $MeV$ and one can
expect that the Big-Bang nucleosynthesis (BBN) took place when the
temperature was not very different from $\sim MeV,$ that is, somewhere in
between seconds and minutes after the Big-Bang. Therefore Big-Bang
nucleosynthesis (BBN), being based on the well understood physics, offers
the possibility of reliable probe of the early universe (see, for instance, 
\cite{W},\cite{KT},\cite{reviews},\cite{OSW} and references cited therein).
The amount of the produced elements depend on the basic cosmological
parameters and is very sensitive to the baryon density. The measured
abundances combined with the CMB temperature fluctuation measurements
provide us an unique opportunity to verify the reliability of the standard
model of the universe evolution \cite{FS},\cite{WMAP}.

The element abundances, are usually calculated using computer codes (for
instance, publically available \cite{sar} Wagoner code \cite{wag}) and the
abundances are presented as the function of the baryon density. To
understand the dependence of the element abundances on the cosmological
parameters the semi-analytical and analytical description of BBN proved to
be very useful \cite{BBr},\cite{Star}. In this paper I develop simple
quasi-equilibrium analytical approach which allows to derive the final
abundances of all light elements up to Beryllium-7 without using any
computer codes. The accuracy of the results is very good for $^{4}He$, good
for $D$ and reasonably good for the other elements. I obtain analytical (not
fitting) formulae describing the dependence of the abundances on the
cosmological parameters and trace the time evolution of the element
abundances before their freeze-out. This highlights an interesting and reach
physics taking place during nucleosynthesis and allows to understand the
physical reasons for the dependence of the abundances on parameters without
practicing with computer codes.

\section{Freeze-out}

The amount of the produced helium depends on the availability of the
neutrons at the time when the helium is formed. In turn, the neutron
concentration is determined by the weak interactions which ensure the
chemical equilibrium between the neutrons and protons at very early time.
The weak interactions become inefficient when the temperature drops below
few $MeV.$ Around this time the neutrons chemically decouple from the
protons and after that the ratio of their concentrations \textquotedblright
freeze out\textquotedblright \footnote{%
The above statement is, of course, literally true only if one neglects the
neutron decay.}. The nuclear reactions take place after that. Therefore,
first we need to calculate the \textquotedblright freeze
out\textquotedblright\ concentration of the neutrons.

The main processes responsible for the chemical equilibrium between protons
and neutrons in the early universe are the weak interaction reactions: 
\begin{equation}
n+\nu \rightleftarrows p+e^{-},\text{ \ \ \ \ \ \ \ }n+e^{+}\rightleftarrows
p+\overline{\nu }.  \label{nuclr1}
\end{equation}%
Here $\nu $ always means the electron-neutrino. To calculate the rate of
these reactions one can use the Fermi theory according to which the matrix
element characterizing 4-fermion interaction (\ref{nuclr1}) is equal: 
\begin{equation}
\left\vert \mathcal{M}\right\vert ^{2}=16\left( 1+3g_{A}^{2}\right)
G_{F}^{2}\left( p_{n}\cdot p_{\nu }\right) \left( p_{p}\cdot p_{e}\right) ,
\label{matr}
\end{equation}%
where $G_{F}=\pi \alpha _{w}/\sqrt{2}M_{W}^{2}\simeq 1.17\times
10^{-5}/GeV^{2}$ is the Fermi coupling constant, $g_{A}\simeq 1.26$ is the
correction to the axial vector \textquotedblright weak
charge\textquotedblright\ of the nucleon\footnote{%
This correction accounts for the possibility that the gluons binding quarks
inside the nucleon can split into quark-antiquark pairs, which could give
nonvanishing contribution to the weak coupling.} and $\left( p_{i}\cdot
p_{j}\right) $ are the scalar products of appropriate 4-momenta entering the
vertex\footnote{%
The Fermi constant can be determined with a very good accuracy measuring the
life time of the muon, while $g_{A}$ can be found only if one considers the
interaction involving the nucleons.}. Considering the process $%
a+b\rightarrow c+d$ of type (\ref{nuclr1}), we get the following expression
for the differential cross-section of this interaction: 
\begin{equation}
\frac{d\sigma _{ab}}{d\Omega }=\frac{1}{\left( 8\pi \right) ^{2}}\frac{%
\left\vert \mathcal{M}\right\vert ^{2}}{\left( p_{a}+p_{b}\right) ^{2}}%
\left( \frac{\left( p_{c}\cdot p_{d}\right) ^{2}-m_{c}^{2}m_{d}^{2}}{\left(
p_{a}\cdot p_{b}\right) ^{2}-m_{a}^{2}m_{b}^{2}}\right) ^{1/2},
\label{difcross}
\end{equation}%
where the integration over the phase space of $c,d-$ particles has been
performed. This expression is manifestly Lorentz-invariant and can be used
in any coordinate frame. Note, that the 4-momenta of the produced particles
are related to the 4-momenta of the colliding particles via the conservation
laws: $p_{c}+p_{d}=p_{a}+p_{b}.$ Let is now consider the particular reaction 
$n+\nu \rightarrow p+e^{-}$ at the temperatures around few $MeV$ and below.
In such a case the nucleons are nonrelativistic; hence 
\begin{equation}
\begin{array}{c}
\left( p_{n}+p_{\nu }\right) ^{2}\simeq m_{n}^{2};\text{ \ \ \ \ }\left(
p_{n}\cdot p_{\nu }\right) =m_{p}\epsilon _{\nu }; \\ 
\sqrt{\left( p_{p}\cdot p_{e}\right) ^{2}-m_{p}^{2}m_{e}^{2}}\simeq m_{p}%
\sqrt{1-\left( m_{e}/\epsilon _{e}\right) ^{2}}=m_{p}\epsilon _{e}v_{e}%
\end{array}
\label{app}
\end{equation}%
where $\epsilon _{\nu }$ is the energy of the incoming neutrino and $%
\epsilon _{e}\simeq \epsilon _{\nu }+Q$ is the energy of the outgoing
electron. The energy $Q=1.293$ $MeV,$ is released when the neutron ``is
converted" into proton. The formula (\ref{difcross}) is directly applicable
only in empty space. However, at the temperatures above $0.5$ $MeV$ there
still present many $e^{\pm }-$pairs and the possible final states for the
electron are partially occupied. Because of the Pauli exclusion principle it
reduces the appropriate cross-section by the factor $\left( 1-n_{\epsilon
_{e}}\right) =\left( 1+\exp \left( -\epsilon _{e}/T\right) \right) ^{-1}.$
Taking this into account and substituting (\ref{app}) into (\ref{matr}), (%
\ref{difcross}) one gets$:$%
\begin{equation}
\sigma _{n\nu }\simeq \frac{1+3g_{A}^{2}}{\pi }G_{F}^{2}\epsilon
_{e}^{2}v_{e}\left( 1+\exp \left( -\epsilon _{e}/T\right) \right) ^{-1}
\label{sigmnnu}
\end{equation}%
where we have neglected the chemical potential of the electrons. Note that
the concentration of the nucleons is negligible compared to the
concentration of the light particles at this time and therefore the spectrum
of the light particles is practically not influenced by the above reactions.
The $n\nu -$interactions taking place within time interval $\Delta t$ in a
given comoving volume, containing $N_{n}$ neutrons, reduce their total
number by amount 
\begin{equation}
\Delta N_{n}=-\left( \sum_{\epsilon _{\nu }}\sigma _{n\nu }n_{\epsilon _{\nu
}}v_{\nu }\Delta g_{\epsilon _{\nu }}\right) N_{n}\Delta t,
\label{intnumber}
\end{equation}%
where $n_{\epsilon _{\nu }}=\left( 1+\exp \left( -\epsilon _{\nu }/T_{\nu
}\right) \right) ^{-1}$, $v_{\nu }=1$ is the speed of the neutrinos and $%
\Delta g_{\epsilon _{\nu }}$ is the phase volume element:%
\[
\Delta g_{\epsilon _{\nu }}=\frac{1}{2\pi ^{2}}\int_{\epsilon _{\nu
}}^{\epsilon _{\nu }+\Delta \epsilon _{\nu }}\left\vert \mathbf{p}%
\right\vert ^{2}d\left\vert \mathbf{p}\right\vert \simeq \frac{1}{2\pi ^{2}}%
\sqrt{\left( \epsilon _{\nu }^{2}-m^{2}\right) }\epsilon _{\nu }\Delta 
\mathbf{\epsilon }_{\nu } 
\]%
Introducing the relative concentration of the neutrons 
\begin{equation}
X_{n}=\frac{N_{n}}{N_{n}+N_{p}}=\frac{n_{n}}{n_{n}+n_{p}},  \label{conn}
\end{equation}%
and substituting (\ref{sigmnnu}) in (\ref{intnumber}) we finally obtain the
following expression for the rate of change of the neutron concentration due
to $n\nu -$ processes 
\begin{equation}
\left( \frac{dX_{n}}{dt}\right) _{n\nu }=-\lambda _{n\nu }X_{n}=-\frac{%
1+3g_{A}^{2}}{2\pi ^{3}}G_{F}^{2}Q^{5}J\left( 1;\infty \right) X_{n},
\label{r1}
\end{equation}%
where 
\begin{equation}
J\left( 1;\infty \right) =\int_{1}^{\infty }dqq^{2}\left( q-1\right)
^{2}\left( 1-\frac{\left( m_{e}/Q\right) ^{2}}{q^{2}}\right) ^{1/2}\left[
1+e^{\frac{Q}{T_{\nu }}\left( q-1\right) }\right] ^{-1}\left[ 1+e^{-\frac{Q}{%
T}q}\right] ^{-1}  \label{r2}
\end{equation}%
and we have introduced the integration variable $q=\left( \epsilon _{\nu
}/Q\right) +1=\epsilon _{e}/Q.$ These expression is given in \cite{W}. If we
neglect the last multiplier into the integrand\footnote{%
This means that one ignores the Pauli's exclusion principle.} and, taking
into account that $q>1$ and $\left( m_{e}/Q\right) ^{2}\simeq \allowbreak
0.15,$ expand the square root keeping only first two terms, the obtained
integral can be calculated exactly and the result is 
\begin{equation}
J\left( 1;\infty \right) \simeq \frac{45\zeta \left( 5\right) }{2}y^{5}+%
\frac{7\pi ^{4}}{60}y^{4}+\frac{3\zeta \left( 3\right) }{2}\left( 1-\frac{1}{%
2}\left( \frac{m_{e}}{Q}\right) ^{2}\right) y^{3}  \label{app1}
\end{equation}%
where $y=T_{\nu }/Q.$ It is quite remarkable that this approximate
expression reproduces the exact result with very high accuracy at all
temperatures. For instance, at $y>1$ the accuracy is about $2\%,$ improving
to $1\%$ and much better for $y<1.$

It is not difficult to understand why it is the case. Actually at low
temperatures $\left( y\ll 1\right) $ this should be so since\footnote{%
We remind that before $e^{\pm }-$annihilation $T=T_{\nu }$ and after that $%
T=1.4T_{\nu }.$} $\exp \left( -Q/T\right) \ll 1.$ On the other hand in the
limit of very high temperatures $\left( y\gg 1\right) $ the integral (\ref%
{r2}) can be very well approximated if one neglects $\left( m_{e}/Q\right) $
and $\left( Q/T_{\nu }\right) -$terms; the result is 
\begin{equation}
J\left( 1;\infty \right) \simeq \frac{7\pi ^{4}}{30}y^{5}\text{ \ \ \ \ at \
\ \ }y\gg 1  \label{app1a}
\end{equation}%
Comparing this with the first term in (\ref{app1}), which obviously
dominates in this limit, we see that they coincide within $3\%-$accuracy
since $\left( 45\zeta \left( 5\right) /2\right) :\left( 7\pi ^{4}/30\right)
=\allowbreak 1.\,\allowbreak 027.$ One can check numerically that in the
intermediate range the accuracy of the approximate expression (\ref{app1})
is better than $2\%;$ for instance, at $y=0.7$ it is about $1\%.$

Substituting (\ref{app1}) together with the numerical values of $G_{F},$ $Q,$
expressed first in the Planck's units, into (\ref{r1}) and then returning
back to the usual units $\left( \left[ \lambda \right] =sec^{-1}\right) $ we
infer that 
\begin{equation}
\lambda _{n\nu }\simeq 1.\,\allowbreak 63y^{3}\left( y+\allowbreak
0.25\right) ^{2}\text{ }sec^{-1},  \label{app2}
\end{equation}%
In this last expression the further simplifications were made. However, the
reader can check himself that at all temperatures $T_{\nu }\geq 0.2$ $MeV$
its accuracy is never worse than $2-3\%$ $.$ Taking into account the
experimental uncertainties in $g_{A}$ this accuracy looks very satisfactory.

\bigskip Similar by, we find that the rate of the reaction $%
n+e^{+}\rightarrow p+\overline{\nu }$ is equal to%
\begin{equation}
\lambda _{ne}=\frac{1+3g_{A}^{2}}{2\pi ^{3}}G_{F}^{2}Q^{5}J\left( -\infty ;-%
\frac{m_{e}}{Q}\right) ,  \label{rne}
\end{equation}%
where $J$ is the integral defined in (\ref{r2}) with the limits of
integration from $-\infty $ to $-\left( m_{e}/Q\right) $. If $T_{\nu }=T$
and $T>m_{e},$ then $\lambda _{ne}\simeq \lambda _{n\nu }.\,$

The rates of the inverse reactions: $pe^{-}\rightarrow n\nu $ and $p%
\overline{\nu }\rightarrow ne^{+}$ are related to the rate of the direct
reactions $\left( \text{at }T_{\nu }=T\right) $ as 
\begin{equation}
\lambda _{pe}=\exp \left( -Q/T\right) \lambda _{n\nu },\text{ \ }\lambda
_{p\nu }=\exp \left( -Q/T\right) \lambda _{ne},\text{\ \ \ }  \label{invrate}
\end{equation}

\bigskip

\textit{\textquotedblright Freeze-out\textquotedblright . }The inverse
reactions lead to the increase of the neutron concentration with the rate $%
\lambda _{p\rightarrow n}X_{p};$ hence we can write the following balance
equation for $X_{n}$: 
\begin{equation}
\frac{dX_{n}}{dt}=-\lambda _{n\rightarrow p}X_{n}+\lambda _{p\rightarrow
n}X_{p}=-\lambda _{n\rightarrow p}\left( 1+e^{-\frac{Q}{T}}\right) \left(
X_{n}-X_{n}^{eq}\right)  \label{eqXn}
\end{equation}%
where $\lambda _{n\rightarrow p}=\lambda _{ne}+\lambda _{n\nu }$ is the
total rate of the direct reactions (\ref{nuclr1}) and $X_{n}^{eq}=\left(
1+\exp \left( Q/T\right) \right) ^{-1}.$ In deriving (\ref{eqXn}) I took
into account that the proton concentration $X_{p}=1-X_{n}$ and used the
relations (\ref{invrate}) assuming that $T_{\nu }=T.$

\bigskip The exact solution of this linear differential equation, with the
initial condition $X_{n}\rightarrow X_{n}^{eq}$ as $t\rightarrow 0,$ can be
written in the following form 
\begin{equation}
X_{n}\left( t\right) =X_{n}^{eq}\left( t\right) -\int_{0}^{t}d\widetilde{t}%
\dot{X}_{n}^{eq}\exp \left( -\int_{\widetilde{t}}^{t}\lambda _{n\rightarrow
p}\left( 1+e^{-\frac{Q}{T}}\right) d\overline{t}\right)   \label{nsol}
\end{equation}%
where dot denotes the derivative with respect to time$.$ At small $t$ the
second term in this equation, characterizing the deviations from the
equilibrium, is negligible compared to the first one. Integrating by parts,
one gets that in this limit the solution (\ref{nsol}) can be rewritten as an
asymptotic series in terms of the derivatives of $X_{n}^{eq}$ 
\begin{equation}
X_{n}=X_{n}^{eq}\left( 1-\frac{1}{\lambda _{n\rightarrow p}}\left( 1+e^{-%
\frac{Q}{T}}\right) ^{-1}\frac{\dot{X}_{n}^{eq}}{X_{n}^{eq}}+....\right) 
\label{nsol1a}
\end{equation}%
Therefore, if $\left\vert \dot{X}_{n}^{eq}/X_{n}^{eq}\right\vert \sim
t^{-1}\ll \lambda _{n\rightarrow p},$ that is the rate of the reactions is
very high compared to the inverse cosmological time, $X_{n}=X_{n}^{eq},$ in
complete agreement with the thermodynamical result. Much later, when the
temperature significantly drops the \textquotedblleft equilibrium
concentration term" $X_{n}^{eq}$ goes to zero and at the same time the
integral on the right hand side of (\ref{nsol}) approaches the finite limit.
As a result the neutron concentration, instead of vanishing, as it would be
in the case of chemical equilibrium, freeze-out at some value $X_{n}^{\ast
}=X_{n}\left( t\rightarrow \infty \right) .$\footnote{%
Note that if $\lambda $ would be decreasing not so fast, such that the
integral in the exponent would diverge as $t\rightarrow \infty ,$ then the
overall integral term would also vanish in this limit.} The freeze-out
effectively occurs when the second term in (\ref{nsol1a}) is of the order of
the first one, that is, when the deviations from the equilibrium become
significant. Assuming that this happens before $e^{\pm }-$ annihilation and
after temperature drops below $Q\simeq 1.29$ $MeV$ (these assumptions can be
checked \textit{a posteriori}) one can put $\lambda _{n\rightarrow p}\simeq $
$2\lambda _{n\nu }$ and neglect $\exp \left( -Q/T\right) $ in the obtained
expressions. In this case the condition $\left\vert \dot{X}%
_{n}^{eq}/X_{n}^{eq}\right\vert \simeq \lambda _{n\rightarrow p},$ defining
the freeze-out temperature $T_{\ast },$ takes the form 
\begin{equation}
y_{\ast }^{2}\left( y_{\ast }+0.25\right) ^{2}\simeq 0.18\kappa ^{1/2}
\label{dev}
\end{equation}%
where $y_{\ast }=T_{\ast }/Q.$ In deriving (\ref{dev}) I used the formula (%
\ref{app2}) for $\lambda _{n\nu }$ and took into account the relation
between the temperature and cosmological time:%
\begin{equation}
t_{sec}=t_{Pl}\left( \frac{3}{32\pi \kappa }\right) ^{1/2}\left( \frac{T_{Pl}%
}{T}\right) ^{2}\simeq 1.39\kappa ^{-1/2}\frac{1}{T_{MeV}^{2}}  \label{cost}
\end{equation}%
where $\kappa \equiv \frac{\pi ^{2}}{30}\left( g_{b}+\frac{7}{8}g_{f}\right) 
$ and $g_{b}$, $g_{f}$ are the total numbers of the internal degrees of
freedom, respectively, of all relativistic bosons and fermions.

In the case of three types of neutrino $(\kappa \simeq 3.54)$ $y_{\ast
}\simeq 0.65$ and the freeze-out temperature is $T_{\ast }\simeq 0.84$ $MeV.$
The equilibrium neutron concentration at this moment is $X_{n}^{eq}\left(
T_{\ast }\right) \simeq 0.18.$ Of course, this number gives only very rough
idea about expected freeze-out concentration. One should not forget that at
this moment the deviations from equilibrium are already very big and, in
fact, $X_{n}\left( T_{\ast }\right) $ exceed the equilibrium concentration
at least twice. The most important thing which could be learned from this
simple estimate is that the freeze-out temperature depends on the number of
light species present in the universe at this time. Since $T_{\ast }\propto
\kappa ^{1/8},$ the more light species are present, the bigger is the
freeze-out temperature and one can expect that more neutrons will survive
after chemical decoupling from the protons\footnote{%
If we use for freeze-out the simple criteria $t\simeq 1/\lambda $ then we
get $T_{\ast }\propto \kappa ^{1/6},$ the result which is usually quoted in
the literature.}. In turn, later on nearly all these neutrons build $^{4}He;$
hence one can expect that if, for instance, in addition to known types of
neutrino, there exist the other light particles, then the abundance of the
primordial helium should be higher than in the case of three neutrinos. This
can be easily understood if we take into account that the rate of the
expansion of the universe $(H=1/2t)$ at the given temperature increases if
we have extra light particles (see (\ref{cost})); hence the freeze-out
should occur earlier, when the neutron concentration is higher. For
instance, in the extreme case of very big number of unknown light particles $%
T_{\ast }\gg Q$ and the expected concentration of the survived neutrons
should be close to $50\%,$ that is, there is one neutron per every proton.
Later on these neutrons would bind the protons converting nearly all
baryonic matter into $^{4}He.$ Of course, this would be in obvious conflict
with the observational abundances of the light elements. Therefore, as we
will see later, the primordial nucleosynthesis allows us to put rather
strong restrictions of the number of light species.

Now I calculate the freeze-out concentration more accurately. Since $%
X_{n}^{eq}\rightarrow 0$ as $T\rightarrow 0,$ this concentration is given by
the integral term in (\ref{nsol}) where we have to take the limit $%
t\rightarrow \infty .$ Changing the integration variable from $t$ to $y=T/Q$
(see (\ref{cost})) and taking into account that the main contribution to the
integral comes at $T>m_{e},$ when $\lambda _{n\rightarrow p}\simeq $ $%
2\lambda _{n\nu }$ and $\lambda _{n\nu }$ is given by (\ref{app2}), we
obtain 
\begin{equation}
X_{n}^{\ast }=\int_{0}^{\infty }\frac{dy}{2y^{2}\left( 1+\cosh \left(
1/y\right) \right) }\exp \left( -5.42\kappa ^{-1/2}\int_{0}^{y}dx\left(
x+0.25\right) ^{2}\left( 1+e^{-1/x}\right) \right)  \label{frout}
\end{equation}%
For the case of three neutrinos $(\kappa \simeq 3.54)$ we get $X_{n}^{\ast
}\simeq 0.158.$ It is in a very good agreement with the results of more
elaborated numerical calculations. The presence of extra light neutrino
increases $\kappa $ by $2\cdot \Delta \kappa _{f}\simeq \allowbreak 0.58$
and respectively the freeze-out concentration becomes $X_{n}^{\ast }\simeq
0.163,$ Hence, two extra fermionic degrees of freedom (one for neutrino and
one for antineutrino) lead to the increase of the freeze-out concentration
by $0.5\%.$

\textit{Neutron decay. }In the above consideration I have neglected the
instability of the neutron via decay 
\begin{equation}
n\rightarrow p+e^{-}+\overline{\nu }  \label{ndecay}
\end{equation}%
It was justified since the lifetime of free neutron $\tau _{n}=885.7\pm 0.8$ 
$sec$ is rather large compared to the typical cosmological time at the
moment of freeze-out $\left( t_{\ast }\sim O\left( 1\right) \text{ }%
sec\right) $. However, later on the two-body reactions (\ref{nuclr1}) and
inverse three-body reaction (\ref{ndecay}) become unimportant and the only
remaining reaction reducing the amount of the neutrons is the neutron decay.
As a result the relative concentration of the neutrons at $t\gg t_{\ast }$
is 
\begin{equation}
X_{n}\left( t\right) =X_{n}^{\ast }\exp \left( -t/\tau _{n}\right)
\label{ndecay1}
\end{equation}%
Note that at late times one can neglect the degeneracy of the leptons which
would increase the lifetime of the free neutrons; hence we can use the
measured in the laboratory lifetime of the neutron quoted above$.$ As we
will see later the nucleosynthesis, as a result of which nearly all neutrons
are captured in the nuclei, where they become stable, happens around $t\sim
200$ $sec.$ It is a rather substantial fraction of the neutron lifetime and
therefore the neutron decay changes significantly the amount of the survived
neutrons and is important for the final $^{4}He-$abundance.

\section{Deuterium bottleneck\textit{\ }}

Complex nuclei are formed as a result of nuclear interactions of the
baryons. For instance, $^{4}He$ could, in principle, be directly formed in
many-body collisions; $2p+2n\rightarrow ^{4}He.$ However, the number
densities at the time when this reaction can take place are too low and its
rate is negligible compared to the rate of expansion. Hence, the light
complex nuclei can be produced only in sequence of two-body reactions. The
first step on this way is the deuterium $\left( D\right) $ production: 
\begin{equation}
p+n\rightleftarrows D+\gamma  \label{deutreac}
\end{equation}%
\ \ There is no problem with this step since the rate of this reaction is
very high and the \textquotedblright typical collision
time\textquotedblright\ is, for sure, much smaller than the cosmological
time (at $t<10^{3}$ $sec$)$.$ Hence one can expect that the deuterium should
be in the local chemical equilibrium with nucleons. Let us define the
deuterium abundance by weight as $X_{D}\equiv 2n_{D}/n_{B,}$ where $n_{B}$
is the total number of all baryons (nucleons) including those ones entering
the complex nuclei. In the state of local chemical equilibrium the relation
between $X_{D}$ and the abundances of the free neutron and protons
(appropriately, $X_{n}\equiv n_{n}/n_{B}$ and $X_{p}\equiv n_{p}/n_{B}$) can
be easily found with the help of the equilibrium Saha's formula (see, for
instance, \cite{KT}):

\begin{equation}
X_{D}=5.67\times 10^{-14}\eta _{10}T_{MeV}^{3/2}\exp \left( \frac{B_{D}}{T}%
\right) X_{p}X_{n}  \label{deut1}
\end{equation}%
where $B_{D}\equiv m_{p}+m_{n}-m_{D}\simeq 2.23$ $MeV$ is the binding energy
of the deuterium and the temperature is expressed in $MeV$. We have
introduced here the normalized baryon-to-photon ratio 
\begin{equation}
\eta _{10}\equiv \eta /10^{-10}=\left( \frac{n_{B}}{n_{\gamma }}\right)
/10^{-10},  \label{bpratio}
\end{equation}%
which is related to the baryon contribution to the critical energy density $%
\Omega _{B}$ as%
\begin{equation}
\Omega _{B}h_{75}^{2}\simeq 6.53\times 10^{-3}\eta _{10}.  \label{bcont}
\end{equation}%
where the Hubble constant $h_{75}$ is normalized on $75$ $km/sec\cdot Mpc$.
The abundance of \ deuterium at the temperatures about its binding energy is
still very small. For instance, for $T\sim 0.5$ $MeV,$ we get $X_{D}\sim
2\times 10^{-13}.$ The reason for that is a very high entropy (number of
photons) per baryon. Even at $T\ll B_{D}$ there are still enough highly
energetic photons with $\epsilon >B_{D}$ which destroy the deuterium.
Actually the number of these photons per one nuclei of the deuterium is
about 
\begin{equation}
\frac{n_{\gamma }\left( \epsilon >B_{D}\right) }{n_{D}}\sim \frac{%
B_{D}^{2}Te^{-B_{D}/T}}{n_{B}X_{D}}\sim 10^{10}\frac{1}{\eta _{10}X_{D}}%
\left( \frac{B_{D}}{T}\right) ^{2}e^{-B_{D}/T}  \label{ng/nd}
\end{equation}%
This number drops below unity at $T<0.06$ $MeV$. Hence one can expect that
the deuterium can be formed in significant amount only when the temperature
is low enough$,$ otherwise it is destroyed by the energetic photons. This
also delays the formation of the other light elements as, for instance, $%
^{4}He.$

The binding energy of the helium-4 $\left( 28.3\text{ }MeV\right) $ is much
higher that the binding energy of the deuterium; hence if helium would be in
chemical equilibrium with neutrons and protons then one would expect that
nearly all free neutrons would be captured in $^{4}He$ already at the
temperature $\sim 0.3$ $MeV.$ However in reality the helium abundance is
still negligible at this temperature. This is because the rates of the
reactions converting deuterium in more heavy elements is proportional to the
deuterium concentration and is much smaller than the expansion rate until
the deuterium abundance will increase and constitute the substantial
fraction of the baryonic matter. Before that only the protons, neutrons and
deuterium are in chemical equilibrium with each other. More heavy elements
are decoupled and present in completely negligible amounts in spite of their
high binding energies. This is known as \textquotedblright deuterium
bottleneck\textquotedblright . The size of the \textquotedblright
bottleneck\textquotedblright\ which is proportional to $X_{D}$ should become
big enough to allow the neutrons and protons \textquotedblright to go
through \textquotedblright\ and replenish the helium abundance in accordance
with its high \textquotedblright equilibrium demand\textquotedblright . Let
us find when this happens. Using formula (\ref{deut1}) we can express the
temperature as a function of $X_{D}$: 
\begin{equation}
T_{MeV}\left( X_{D}\right) \simeq \frac{0.061}{\left( 1+2.7\times 10^{-2}\ln
\left( X_{D}/\eta _{10}\right) \right) }  \label{tx}
\end{equation}%
This relation is valid only when the deuterium is in chemical equilibrium
with neutrons and protons, which as we will see is true until the moment
when $X_{D}$ reaches the value $10^{-2}.$ According to the formula (\ref{tx}%
) the deuterium abundance should change from $10^{-5}$ to $1$ when the
temperature drops only in $1.5$ times, namely, from $0.09$ $MeV$ to $0.06$ $%
MeV$ (for $\eta _{10}=1$). Therefore, the deuterium abundance should
increase very abruptly around this time and one can expect that the nuclear
reactions should become fast enough to proceed with formation of the light
elements. The main processes converting the deuterium in more heavy elements
are (see also Fig.1): 
\begin{equation}
1)\text{ }D+D\rightarrow ^{3}He+n,\text{ \ \ \ \ }2)\text{ }D+D\rightarrow
T+p  \label{DD}
\end{equation}%
The cross-sections of these reactions are known from experiments and the
results are usually presented in terms of the effective rates vs. temperature%
\footnote{%
The appropriate rates are cited in \cite{Star}. More recent data can be
found on internet.}. In the temperature interval $0.06\div 0.09$ $MeV$ these
rates change not very much and we have 
\begin{equation}
\left\langle \sigma v\right\rangle _{DD1}=\left( 1.3\div 2.2\right) \times
10^{-17}cm^{3}/sec,\text{ \ \ \ \ }\left\langle \sigma v\right\rangle
_{DD2}=\left( 1.2\div 2\right) \times 10^{-17}cm^{3}/sec\text{\ \ \ }
\label{rdd}
\end{equation}%
Considering the comoving volume containing $N_{D}$ deuterium nuclei we find
that the decrease of their number during the time interval $\Delta t$ due to
the reactions (\ref{DD}) is equal to 
\begin{equation}
\Delta N_{D}=-\left\langle \sigma v\right\rangle _{DD}n_{D}N_{D}\Delta t
\label{Deq}
\end{equation}%
Rewriting this equation in terms of the concentration by weight $X_{D}\equiv
2N_{D}/N_{B}$ we obtain 
\begin{equation}
\Delta X_{D}=-\frac{1}{2}\lambda _{DD}X_{D}^{2}\Delta t  \label{deq1}
\end{equation}%
where 
\begin{equation}
\lambda _{DD}=\left( \left\langle \sigma v\right\rangle _{DD1}+\left\langle
\sigma v\right\rangle _{DD2}\right) n_{B}\sim 1.3\times 10^{5}K\left(
T\right) T_{MeV}^{3}\eta _{10}\text{ }sec^{-1}  \label{r11}
\end{equation}%
and $K\left( T\right) $ is the numerical coefficient which changes from $%
\simeq 1$ to $\simeq 0.6$ when the temperature drops from $0.09$ $MeV$ to $%
0.06$ $MeV.$ It is clear that the substantial amount of the available
deuterium can be converted into helium-3 and tritium within the cosmological
time $t$ only if $\Delta X_{D}\simeq \left( 1/2\right) \lambda
_{DD}X_{D}^{2}t\sim X_{D};$ hence the \textquotedblright deuterium
bottleneck opens wide\textquotedblright\ only when the deuterium
concentration reaches the value 
\begin{equation}
X_{D}^{\left( i\right) }\simeq \frac{1.2\times 10^{-5}}{\eta
_{10}T_{MeV}\left( X_{D}\right) }\simeq 1.5\times 10^{-4}\eta
_{10}^{-1}\left( 1-7\times 10^{-2}\ln \eta _{10}\right)   \label{begnucl}
\end{equation}%
Deriving this formula I used the relations (\ref{cost}) with $\kappa \simeq
1.11,$ and (\ref{tx}); the obtained equation was solved by iterations
assuming that $10^{-1}<\eta _{10}<10.$

After deuterium abundance reaches the value given by (\ref{begnucl})
everything proceeds very fast. In fact, if $\eta _{10}=1$ then according to (%
\ref{tx}) the equilibrium concentration $X_{D}$ should increase from $%
10^{-4} $ to $10^{-2}$ when the temperature drops from $0.08$ $MeV$ to $0.07$
$MeV$. This increase of $X_{D}$ means that the reaction rates converting the
deuterium to more heavy elements, which are proportional to $X_{D}^{2},$ at $%
T\sim 0.07$ $MeV$ become $10^{4}$ times bigger than the rate of the
expansion. It is clear that this system is far from the equilibrium and the
deuterium supplied by $pn-$reactions \textquotedblleft is converted" very
fast to more heavy elements. This doesn't allow the deuterium concentration
to increase to the values bigger than $10^{-2}.$ The details of the
nonequilibrium processes are described by a complicated system of kinetic
equations which can be solved only numerically. In Fig.2 the results of
numerical calculations for the time evolution of the element abundances in
the universe with $\Omega _{B}h_{75}^{2}\simeq 5\times 10^{-2}$ are shown 
\cite{BNT}.

Below I present the calculations which explain the time behavior of these
abundances and derive the formulae for the final freeze-out abundances of
light elements up to $^{7}Be.$ This includes $^{4}He,$ deuterium $\left(
D\right) ,$ helium-3 $\left( ^{3}He\right) ,$ tritium $\left( T\right) ,$
Lithium-7 $\left( ^{7}Li\right) $ and Beryllium $\left( ^{7}Be\right) .$ The
other light elements as, for instance, $^{8}Li$, $^{8}B$ etc. are produced
in very small amounts and will be ignored.

\bigskip

The \textit{most important} nuclear reactions involving the light elements
are schematically depicted in Fig.1, which I recommend to keep in front of
the eyes reading the rest of the paper.

\begin{figure}[h]
\begin{center}
\includegraphics[width=13cm,angle=0]{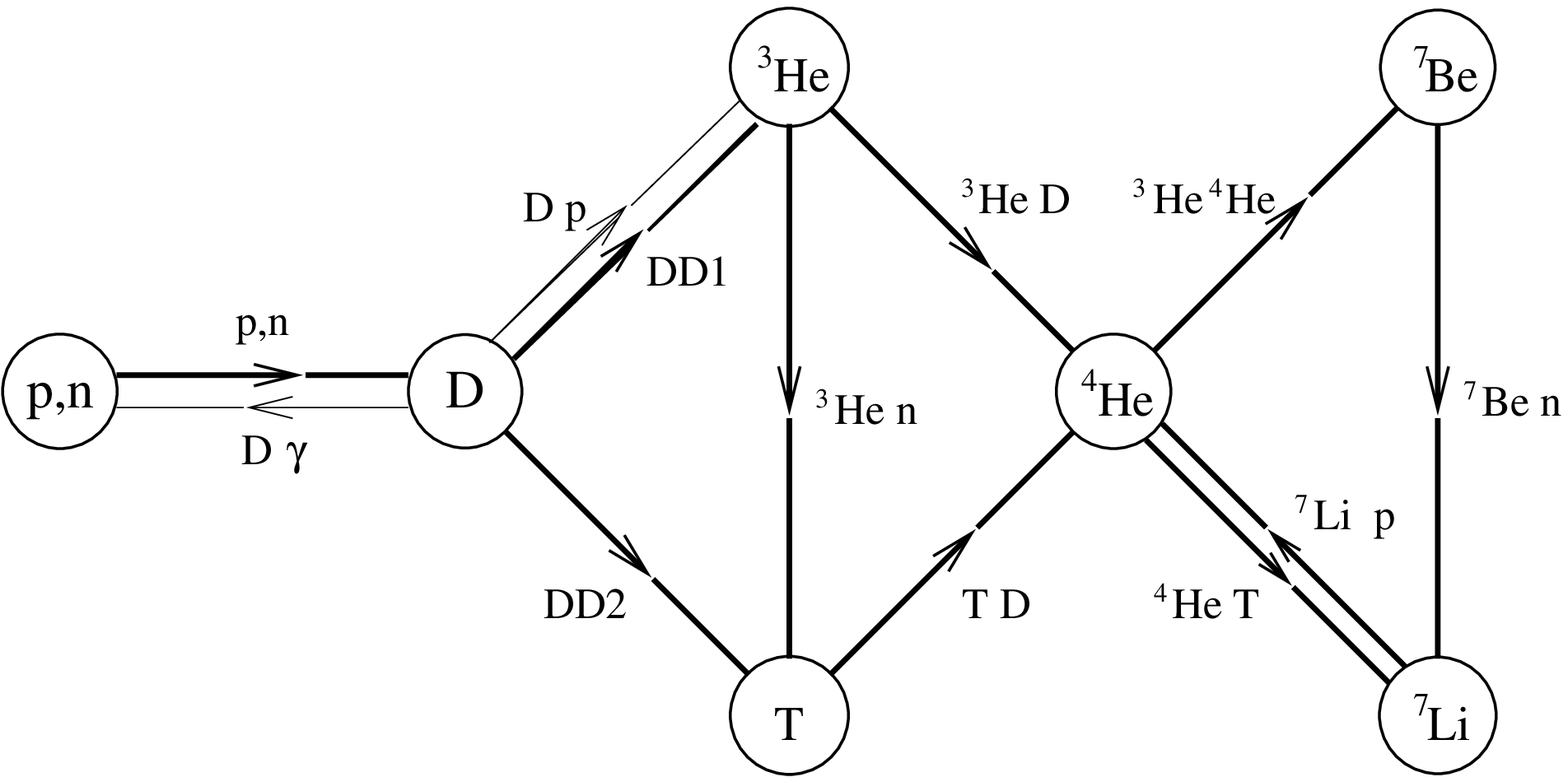} \vspace{0.1cm}\\[0pt]
figure 1
\end{center}
\end{figure}

In this Figure to every element corresponds its own \textquotedblright
reservoir\textquotedblright\ . All these \textquotedblright
reservoirs\textquotedblright\ are connected by \textquotedblright
one-way-pipes\textquotedblright . Every \textquotedblright
pipe\textquotedblright\ corresponds to an appropriate nuclear reaction. I
write only the initial elements involved in the reaction, since the outcome
can be easy inferred from the picture. The \textquotedblright thickness of
the pipe\textquotedblright\ through which the element $a$ \textquotedblright
escape from the reservoir\textquotedblright\ as a result of the reaction $%
ab\rightarrow cd$ is proportional to rate of this reaction 
\begin{equation}
\dot{X}_{a}/X_{a}=-A_{b}^{-1}\lambda _{ab}X_{b}
\end{equation}%
where $\lambda _{ab}=\left\langle \sigma v\right\rangle _{ab}n_{B}$ and $%
A_{b}$ is the mass number of the element $b;$ for instance, $\ A=4$ for $%
^{4}He$ and $A=7$ for $^{7}Li,$ $^{7}Be.$ Of course, the appropriate
\textquotedblright pipe\textquotedblright\ is efficient only if $\dot{X}%
_{a}/X_{a}>t^{-1}.$ As we have already seen the $D-$ and $p,n-$
reservoirs\textquotedblright\ are in equilibrium with each other and
decoupled from the rest at the temperatures above $0.08$ $MeV$
(\textquotedblright deuterium bottleneck\textquotedblright ). However when
the temperature drops below $0.08$ $MeV$ the \textquotedblright $DD-$pipes
open\textquotedblright\ and become very efficient in converting an extra
deuterium supply from \textquotedblright $np-$reservoir\textquotedblright\
into more heavy element. Finally nearly all free neutrons disappear entering
more heavy elements where they become stable. After that the concentrations
of the elements in the appropriate \textquotedblright
reservoirs\textquotedblright\ freeze-out and the ``final abundances"
survive. This is a general picture and now I proceed with detailed
calculations and consider the formation of every element separately.

\section{Helium-4}

As soon as deuterium concentration increases to $X_{D}^{\left( i\right) }$
given by (\ref{begnucl}) the formation of the other light elements begins.
This happens at the temperature (see (\ref{tx})) 
\begin{equation}
T_{MeV}^{\left( i\right) }\simeq 0.08\left( 1+7\times 10^{-2}\ln \eta
_{10}\right)  \label{bntem}
\end{equation}%
at the moment of time\footnote{%
Note that $T^{\left( i\right) }$ and $t^{\left( i\right) }$ depend on the
exact value of $X_{D}^{\left( i\right) }$ only logarithmically and therefore
not very sensitive to the exact value of $X_{D}^{\left( i\right) }.$} 
\begin{equation}
t_{sec}^{\left( i\right) }\simeq 206\left( \frac{\kappa }{1.11}\right)
^{-1/2}\left( 1-0.14\ln \eta _{10}\right) \text{ }  \label{bntime}
\end{equation}%
Of course, the nucleosynthesis does not happen instantaneously. Moreover at
the beginning the rate of deuterium production in reaction, $pn\rightarrow
D\gamma ,$ is substantially higher that the total rate of the deuterium
\textquotedblright annihilation\textquotedblright\ in reactions (\ref{DD}),
namely, 
\begin{equation}
\frac{\lambda _{pn}X_{p}X_{n}}{\lambda _{DD}X_{D}^{2}}\simeq 10^{4}\left( 
\frac{10^{-4}}{X_{D}}\right) ^{2}  \label{pn/DD}
\end{equation}%
where I used the experimental value for the ratio $\lambda _{pn}/\lambda
_{DD},$ which is about $10^{-3}$ at $T_{MeV}\simeq 0.07\div 0.08$ and put $%
X_{n}=1-X_{p}\simeq 0.16.$

As it follows from (\ref{pn/DD}) before the deuterium concentration reaches
its maximal value $X_{D}\sim 10^{-2}$ the deuterium production dominates
over deuterium destruction and the deuterium abundance continues to follow
its chemical equilibrium track given by (\ref{deut1}). According to (\ref{tx}%
) the concentration $X_{D}\simeq 10^{-2}$ is reached very fast after $%
t^{\left( i\right) }$, namely, when the temperature drops from $0.08$ $MeV$
to $0.07$ $MeV$ $\left( \text{for }\eta _{10}=1\right) ,$ that is, with 
\begin{equation}
\Delta t\simeq 2t^{\left( i\right) }\frac{\Delta T}{T^{\left( i\right) }}%
\simeq 50\text{ }sec  \label{del}
\end{equation}%
time delay after $t^{\left( i\right) }.$ When this concentration is reached
the two-body $DD-$deuterium destruction become more efficient than the $pn-$%
deuterium production and $X_{D}$ begins to decrease\footnote{%
The deuterium photo-destruction can be completely neglected after that. It
is clear if we note that if there would be only photo-destruction processes
alone then the deuterium concentration would continue to increase.} (see
Fig.2).

\begin{figure}[h]
\begin{center}
\includegraphics[width=10cm]{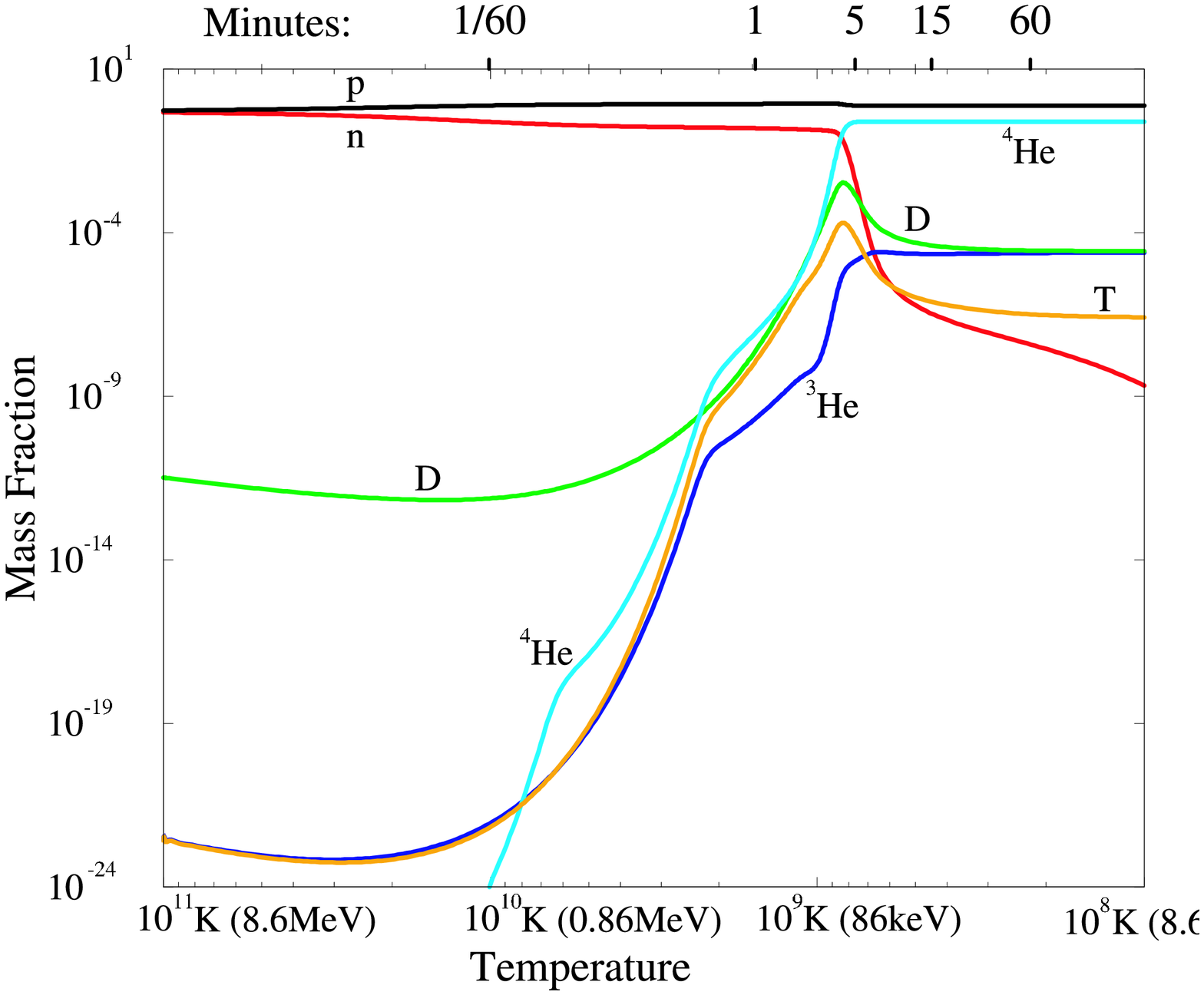} \vspace{0.1cm}\\[0pt]
figure 2
\end{center}
\end{figure}

The concentration of the free neutrons during this period strongly decreases
and they go first to the \textquotedblright deuterium
reservoir\textquotedblright\ and then proceed further \textquotedblright
through the pipes\textquotedblright\ forming heavy elements. For most
neutrons the ``final destination" is the \textquotedblright $^{4}He-$%
reservoir\textquotedblright .

Why it is so can be understood even without analyzing the rates of the
intermediate reactions. Actually, if $^{4}He$ would be in the equilibrium
with the other light elements it would be dominating at low temperatures
because of its high binding energy $\left( 28.3\text{ }MeV\right) ,$ which
is four times bigger than the binding energies of the intermediate elements, 
$^{3}He$ $\left( 7.72\text{ }MeV\right) $ and $T$ $\left( 6.92\text{ }%
MeV\right) $. The system which is away from equilibrium always tends there
in a quickest possible way. Therefore, most of the free neutrons will be
capture into $^{4}He-$nuclei because its equilibrium demand is the highest.

The reactions proceed in the following way. First, the deuterium is
converted into $^{3}He$ and $T$ in reactions (\ref{DD}). After that tritium
interacts with deuterium and produce the helium-4 nuclei: 
\begin{equation}
T+D\rightarrow ^{4}He+n  \label{TD}
\end{equation}%
As a result two neutrons out of three are captured into the $^{4}He-$nuclei
and one comes back into "$np-$reservoir".

The $^{3}He-$ nuclei can interact either with free neutrons and then proceed
to \textquotedblright $T-$reservoir\textquotedblright , 
\begin{equation}
^{3}He+n\rightarrow T+p,  \label{He3n}
\end{equation}%
or with deuterium going directly to \textquotedblright $^{4}He-$%
reservoir\textquotedblright\ 
\begin{equation}
^{3}He+D\rightarrow ^{4}He+p  \label{He3D}
\end{equation}%
The ratio of rates for these reactions is 
\begin{equation}
\frac{\lambda _{^{3}Hen}X_{^{3}He}X_{n}}{\lambda _{^{3}HeD}X_{^{3}He}X_{D}}%
\sim 6\frac{X_{n}}{X_{D}};  \label{ratHe3n}
\end{equation}%
hence at the beginning \textquotedblright $^{3}HeD-$pipe\textquotedblright\
is inefficient compared to \textquotedblright $^{3}Hen-$pipe%
\textquotedblright\ and most of $^{3}He-$ nuclei are converted into tritium.
Only when the concentration of the free neutrons drops below the deuterium
concentration (which is always smaller that $10^{-2}$), the rate of the
reaction (\ref{He3D}) converting $^{3}He$ directly into $^{4}He$ becomes
bigger that the rate of the reaction (\ref{He3n}). It follows from here that
most of the neutrons will go into $^{4}He-$nuclei either along $%
np\rightarrow D\rightarrow T\rightarrow ^{4}He$ or $np\rightarrow
D\rightarrow ^{3}He\rightarrow T\rightarrow ^{4}He$ way$.$ Finally, in about 
$50\div 100$ $sec$ after the beginning of nucleosynthesis nearly all
neutrons (with the exception of very small fraction $<10^{-3}$)$,$ end up in 
$^{4}He-$nuclei. Therefore, the final $^{4}He-$abundance is completely
determined by the amount of the available free neutrons at the time when $DD$%
-reactions become efficient, that is at $t\simeq t^{\left( i\right) }.$
Because half of the total weight of $^{4}He$ is due to the protons, its
final abundance by weight should be 
\begin{equation}
X_{^{4}He}^{f}=2X_{n}\left( t^{\left( i\right) }\right) =2X_{n}^{\ast }\exp
\left( -\frac{t^{\left( i\right) }}{\tau _{n}}\right)   \label{he4-1}
\end{equation}%
Substituting here $X_{n}^{\ast }$ from (\ref{frout}) and $t^{\left( i\right)
}$ from (\ref{bntime}) we obtain: 
\begin{eqnarray}
X_{^{4}He}^{f} &=&2\left( 0.158+0.005\left( N_{v}-3\right) \right) \cdot
\exp \left( -\frac{206}{886}\frac{\left( 1-0.14\ln \eta _{10}\right) }{%
\left( 1+\frac{0.15}{1.11}\left( N_{v}-3\right) \right) ^{1/2}}\right)  
\nonumber \\
&\simeq &0.25+0.012\left( N_{\nu }-3\right) +0.0082\ln \eta _{10}
\label{he42}
\end{eqnarray}%
where $N_{\nu }$ is the number of massless neutrino species. This result is
in a very good agreement with the results of the numerical calculations
presented in Fig.3 \cite{OSW}.

\begin{figure}[h]
\begin{center}
\includegraphics[width=9cm,angle=0]{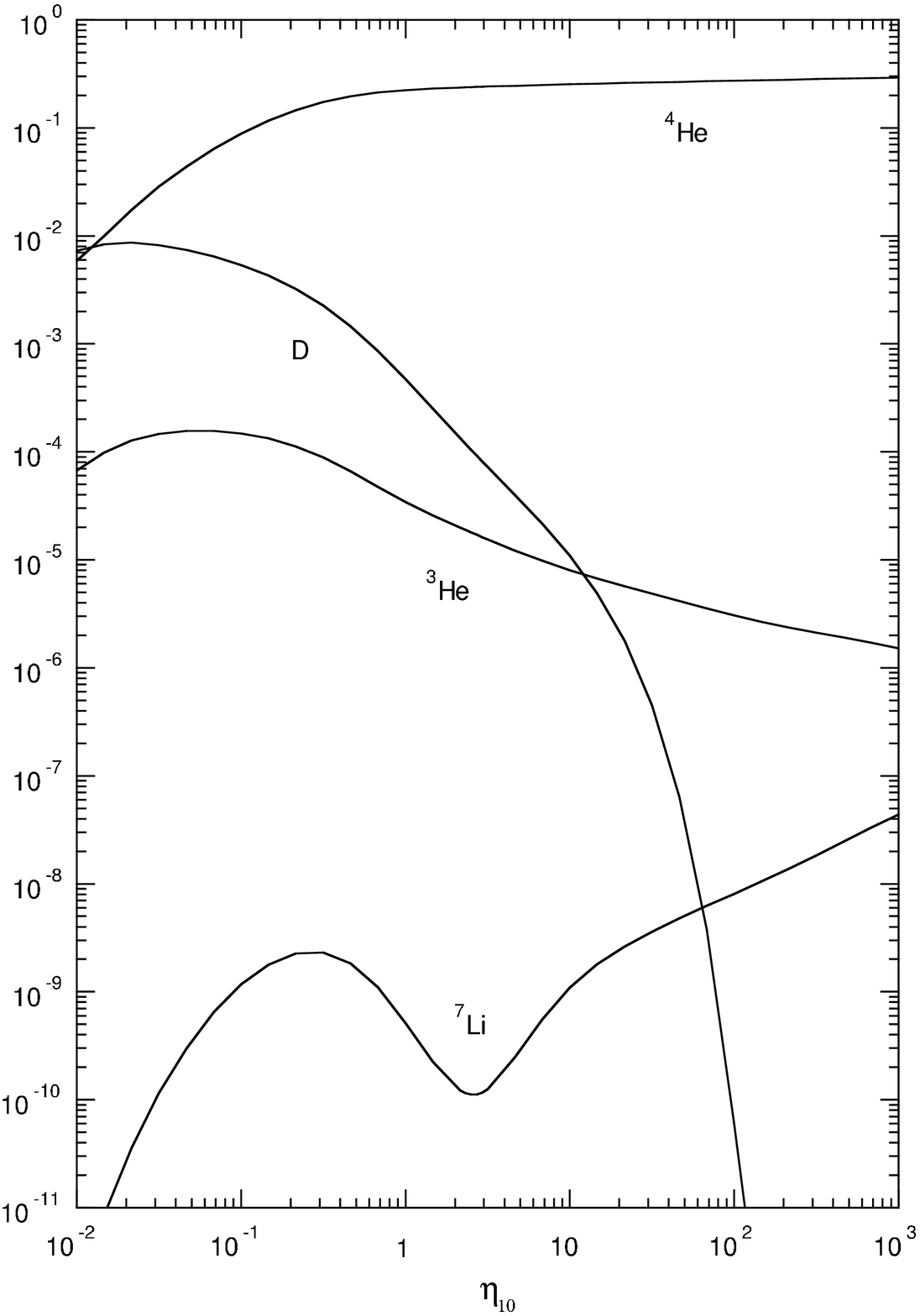} \vspace{0.5cm}\\[0pt]
figure 3
\end{center}
\end{figure}

In fact, this agreement can be made even better if one notes that the
formation of $^{4}He$ is not an instantaneous event which happens at $%
t^{\left( i\right) }.$ It starts at $t^{\left( i\right) }$ and then
continues for, at least, $50$ $sec$ (see (\ref{del})). Most of the neutrons
are trapped at the end. Therefore the time delay reduces the amount of $%
^{4}He$ to $X_{^{4}He}\simeq 0.25\exp \left( -50/886\right) \simeq
\allowbreak 0.236$ that is by $1,4\%.$

As we see from (\ref{he42}) the abundance of $^{4}He$ depends on the number
of massless species $N_{\nu }$. The presence of extra massless neutrino
increases the $^{4}He-$abundance by $1.2\%.$ There are two reasons for this.
Two third out of this increase is due to the dependence of the freeze-out
concentration $X_{n}^{\ast }$ from the number of the massless species. In
fact, more species one has, more fast universe expands at given temperature
and hence the freeze-out of the neutrons occurs earlier, when their
concentration is higher. The remaining one third has a similar nature.
Namely, for given baryon density the nucleosynthesis happens at appropriate
temperature. This temperature is reached earlier if there are more light
species and therefore more neutrons survive until they will be captured. The
dependence of the $^{4}He-$abundance on the number of light species taken
together with the results for the deuterium abundance allows us to put
rather strong bounds on the number of unknown light particles which were
relativistic at the time of nucleosynthesis.

The helium abundance also depends on the baryon density (entropy per baryon)
and according to (\ref{he42}) increases by $\sim $ $2\%$ (numerical result $%
\simeq 2.5\%$) if the density is ten times higher. The physical origin of
this dependence is very clear. In the universe with bigger concentration of
baryons the nucleosynthesis begins earlier, at higher temperature (see (\ref%
{bntem})); hence more neutrons survive till this time and more Helium-4 is
formed.

\section{Deuterium}

To calculate the time evolution and freeze-out concentration of deuterium I
will make some assumption which significantly simplify the consideration.
The validity of these assumption can be checked \textit{a posteriori.}

First of all, I ignore $^{7}Be,$ $^{7}Li$ since their abundances as we will
see later are always small compared to the abundances of $^{3}He$ and $T$.
Second, I will assume \textit{that }$^{3}He$\textit{\ and }$T$\textit{\
abundances} always have \textit{quasi-equilibrium values}, which are
determined by condition that the \textit{\textquotedblright total incoming
in appropriate reservoir flux should be equal to the outgoing
flux\textquotedblright }\footnote{%
This condition reminds the first Kirchhoff's rule for the electric currents.}%
(see Fig.1). For instance, in the case of $^{3}He$ it means that the amount
of $^{3}He$ produced within some time interval in $DD$ and $Dp-$ reactions
should be equal to the amount of $^{3}He$ destroyed during the same time in $%
^{3}HeD$ and $^{3}Hen-$reactions. This is well justified because the rate of
the reactions in which $^{3}He$ is destroyed is high enough to take care
about \textquotedblleft quick adjustment" of $^{3}He-$concentration to the
change of deuterium abundance.

The system of reservoirs with pipes, depicted in Fig.1 is a
\textquotedblright self-regulated system\textquotedblright\ with small
adjustment time. The overall picture after the beginning of the
nucleosynthesis is the following. When deuterium concentration reaches $%
X_{D}\simeq 10^{-2}$ the rate of $DD-$reactions become comparable with the
rate of the deuterium production via $pn-$interactions (see(\ref{ratHe3n}))
and then dominates. The neutrons are taken from \textquotedblright $np-$%
reservoir\textquotedblright\ and send via \textquotedblright $D-$%
reservoir\textquotedblright\ along \textquotedblright $DD$ and $Dp-$%
pipes\textquotedblright\ first to \textquotedblright $^{3}He$ and $T-$%
reservoirs\textquotedblright\ and from there through \textquotedblright $%
^{3}HeD$ and $TD-$pipes\textquotedblright\ to their final destination,
namely, in \textquotedblright $^{4}He-$reservoir\textquotedblright . Not all
of the neutrons taken from \textquotedblright $np-$reservoir%
\textquotedblright\ reach the \textquotedblright $^{4}He-$%
reservoir\textquotedblright\ in the first try. Some of them
\textquotedblright escape\textquotedblright\ on the way there. Namely, in
\textquotedblright $DD1$ and $TD$-pipes\textquotedblright\ one neutron is
released in the reactions $DD\rightarrow ^{3}Hen,$ $TD\rightarrow ^{4}Hen$,
comes back to \textquotedblright $np-$reservoir\textquotedblright\ and then
participate in the next try to get \textquotedblright $^{4}He-$%
reservoir\textquotedblright . Thus after the beginning of nucleosynthesis
there is a stationary flux of the neutrons from \textquotedblright $pn-$%
reservoir\textquotedblright\ to \textquotedblright $^{4}He-$%
reservoir\textquotedblright\ through the system of \textquotedblright
pipes\textquotedblright\ via intermediate \textquotedblright $D,^{3}He$ and $%
T-$reservoirs\textquotedblright . The ``widths of the pipes" (reaction
rates) connected to $^{3}He$ and $T-$reservoirs depend on the concentration
in the appropriate reservoir. For instance, the width of the
\textquotedblright $^{3}HeD$ and $TD-$pipes is proportional, respectively,
to $^{3}He$ and $T-$concentrations. If the amount of $^{3}He$ would
increase/decrease compared to its quasi-equilibrium value the size of
\textquotedblright $^{3}HeD$ $-$pipe\textquotedblright\ would be quickly
adjusted (respectively increases/decreases) to bring its concentration to
the quasi-equilibrium value, which is determined by the condition of \textit{%
zero total flux}.

If the universe would not be expanding then finally nearly all neutrons
would go to \textquotedblright $^{4}He-$reservoir\textquotedblright\ and as
long as the temperature goes to zero, nothing would be left besides of the
protons and $^{4}He.$ However, the expansion plays the role of
\textquotedblright water-tap\textquotedblright\ for the \textquotedblright
pipes\textquotedblright . At the moment when the reaction rates become
smaller than the rate of the expansion the \textquotedblright
water-taps\textquotedblright\ close and the abundances of the elements in
the appropriate reservoirs freeze-out at their quasi-equilibrium values. The
final abundances of $^{3}He$ and $T$ are determined by deuterium freeze-out
concentration which we have to calculate.

Analyzing the system of kinetic equations one can find that even if $^{3}He$
and $T$ have quasi-equilibrium concentrations the neutrons and deuterium
concentrations \textit{not necessary satisfy} the quasi-equilibrium
conditions$.$ Therefore, we have to derive the equations which describe the
time dependence of the appropriate abundances $X_{n}$, $X_{D}$ after $X_{D}$
reached the value $\sim 10^{-2}$.

The reaction rate for the elements $a,b$ which is equal to $\lambda
_{ab}n_{a}n_{b}/n_{B}^{2}$ can be rewritten in terms of the concentrations
by weight as 
\begin{equation}
\frac{1}{A_{a}A_{b}}\lambda _{ab}X_{a}X_{b}  \label{rr}
\end{equation}%
where $A_{a},A_{b}$ are the mass numbers of the elements $a$ and $b.$ The
quasi-equilibrium condition for $^{3}He$ takes then the following form: 
\begin{equation}
\frac{1}{4}\lambda _{DD1}X_{D}^{2}+\frac{1}{2}\lambda _{Dp}X_{D}X_{p}=\frac{1%
}{6}\lambda _{^{3}HeD}X_{^{3}He}X_{D}+\frac{1}{3}\lambda
_{^{3}Hen}X_{^{3}He}X_{n}  \label{qeHe3}
\end{equation}%
Similar for tritium we have 
\begin{equation}
\frac{1}{4}\lambda _{DD2}X_{D}^{2}+\frac{1}{3}\lambda
_{^{3}Hen}X_{^{3}He}X_{n}=\frac{1}{6}\lambda _{TD}X_{T}X_{D}  \label{qeT}
\end{equation}%
I will assume that these conditions are always satisfied.

The general kinetic equation for the rate of change of free neutrons
concentration can be easily written if we take into account they are
produced in the reactions $DD\rightarrow ^{3}Hen$ and $DT\rightarrow $ $%
^{4}Hen$ and \textquotedblright destroyed\textquotedblright\ in the
processes $pn\rightarrow D\gamma $ and $^{3}Hen\rightarrow Tp:$%
\begin{equation}
\frac{dX_{n}}{dt}=\frac{1}{4}\lambda _{DD1}X_{D}^{2}+\frac{1}{6}\lambda
_{TD}X_{T}X_{D}-\lambda _{pn}X_{p}X_{n}-\frac{1}{3}\lambda
_{^{3}Hen}X_{^{3}He}X_{n}.  \label{xneq}
\end{equation}%
Assuming that tritium satisfies quasi-equilibrium condition (\ref{qeT}) one
can simplify this equation to 
\begin{equation}
\frac{dX_{n}}{dt}=\frac{1}{4}\lambda _{DD}X_{D}^{2}-\lambda _{pn}X_{p}X_{n},
\label{xneq1}
\end{equation}%
where as usually $\lambda _{DD}=\lambda _{DD1}+\lambda _{DD2}.)$

The appropriate equation for\textbf{\ }deuterium is derived similar by,
using (\ref{qeHe3}) and (\ref{qeT}): 
\begin{equation}
\frac{dX_{D}}{dt}=2\lambda _{pn}X_{p}X_{n}-\lambda _{DD}X_{D}^{2}-2\lambda
_{Dp}X_{D}X_{p}.  \label{xdeq}
\end{equation}

Expressing time through the temperature via (\ref{cost}) and substituting
the numerical values for $\lambda _{DD}$ given by (\ref{r11}), the above
equations reduce to 
\begin{equation}
\frac{dX_{n}}{dT_{MeV}}=a\cdot K\left( T\right) \eta _{10}\left(
R_{1}X_{n}-X_{D}^{2}\right)  \label{feq1}
\end{equation}%
and 
\begin{equation}
\frac{dX_{D}}{dT_{MeV}}=4a\cdot K\left( T\right) \eta _{10}\left(
X_{D}^{2}+R_{2}X_{D}-\frac{1}{2}R_{1}X_{n}\right)  \label{feq2}
\end{equation}%
where $a=0.86\times 10^{5}$ and the coefficient $K\left( T\right) $ accounts
for the temperature dependence of $\left\langle \sigma v\right\rangle $ for $%
DD$-reactions and changes from $\sim 1$ to $0.5$ when the temperature drops
from $0.09$ $MeV$ to $0.04$ $MeV.$ In the expressions 
\begin{equation}
R_{1}\equiv 4X_{p}\frac{\lambda _{pn}}{\lambda _{DD}}\simeq \left( 3\div
8\right) \times 10^{-3},\text{ \ \ \ }R_{2}\equiv 2X_{p}\frac{\lambda _{pD}}{%
\lambda _{DD}}\simeq \left( 2.5\div 2.3\right) \times 10^{-5}  \label{RR}
\end{equation}%
I used the experimental value for the ratio of the appropriate reaction
rates; the first number within the brackets corresponds to the higher
temperature when it changes in the interval $T_{MeV}\simeq 0.09\div 0.04.$

The system of equations (\ref{feq1}) and (\ref{feq2}) has an attractor
solutions, which can be easily found if we consider $X_{D}$ as a function $%
X_{n}$ (or vise versa) and rewrite the eqs. (\ref{feq1}),(\ref{feq2}) as%
\[
\frac{dX_{D}}{dX_{n}}=4\frac{\left( X_{D}^{2}+R_{2}X_{D}-\frac{1}{2}%
R_{1}X_{n}\right) }{\left( R_{1}X_{n}-X_{D}^{2}\right) }
\]%
If $X_{D}\ll X_{n}$ then 
\begin{equation}
X_{n}=\frac{2}{R_{1}}\left( X_{D}^{2}+R_{2}X_{D}\right) \left[ 1-\frac{1}{8}%
\frac{X_{D}}{X_{n}}+O\left( \left( \frac{X_{D}}{X_{n}}\right) ^{2}\right) %
\right]   \label{sol1a}
\end{equation}%
satisfies this equation up to the second order terms in $X_{D}/X_{n}.$ The
solution (\ref{sol1a}) is a good approximate solution \textit{after}
deuterium concentration reaches the maximal value about $10^{-2}$ and begins
to decrease (see Fig.2). It is valid until the moment when the neutron
concentration drops and becomes comparable to the deuterium concentration.
The solution (\ref{sol1a}) describes the situation when the deuterium
abundance satisfy the quasi-equilibrium condition. One can check that in
this case the time derivative of the deuterium concentration in the l.h.s.
of the equation (\ref{feq2}) is small compared to every separate term in the
r.h.s. of this equation. Since $R_{2}\ll R_{1}$ we infer from the eq. (\ref%
{sol1a}) that the deuterium and neutron concentration become comparable when
the deuterium concentration drops to $O\left( 1\right) R_{1}.$ Before this
happens (for $X_{D},X_{n}>O\left( 1\right) R_{1}$) the deuterium
concentration can be expressed through the neutron concentration as 
\begin{equation}
X_{D}\simeq \sqrt{\frac{R_{1}X_{n}}{2}.}  \label{sold1}
\end{equation}%
Note that according to this formula the maximal possible concentration which
deuterium can reach is $X_{D}\simeq 10^{-2}$ when most the free neutrons are
still not captured by light elements $\left( X_{n}\simeq 0.12\right) .$ This
is in complete agreement with naive estimate we got before comparing $pn-$
and $DD-$ reactions rates. When deuterium follows its quasi-equilibrium
track (\ref{sold1}) the neutrons concentration satisfies the equation 
\begin{equation}
\frac{dX_{n}}{dT_{MeV}}\simeq \frac{1}{2}a\cdot K\left( T\right) \eta
_{10}R_{1}X_{n}  \label{n1}
\end{equation}%
In this case the neutrons are the \textquotedblright key
element\textquotedblright\ which determines the quasi-equilibrium
concentrations of all other elements including deuterium. In other words,
the neutrons regulate the \textquotedblright water-taps in the pipes
connecting the reservoirs in Fig.1\textquotedblright .\ The equation (\ref%
{n1}) starts to be applicable at the moment when deuterium concentration
grows to $10^{-2}.$ At this time most of the free neutrons are not yet
trapped by the light elements and $X_{n}\simeq 0.12.$ According to (\ref{tx}%
), which is still applicable at this time, the deuterium reaches the maximal
possible concentration $\sim 10^{-2}$ when the temperature drops to%
\begin{equation}
T_{MeV}\simeq 0.07+0.002\ln \eta _{10}
\end{equation}%
After that the neutron concentration satisfies the equation (\ref{n1}), the
approximate solution of which is 
\begin{equation}
X_{n}\left( T_{MeV}\right) \simeq 0.12\exp \left( \frac{1}{2}a\cdot K\left(
T\right) \eta _{10}R_{1}\left( T_{MeV}-0.07-0.002\ln \eta _{10}\right)
\right) 
\end{equation}%
As it follows from here, the neutron concentration decreases as the
temperature drops and becomes equal to the deuterium concentration $\left(
\sim R_{1}\right) $ when 
\begin{equation}
T_{MeV}^{\ast }\sim 0.07-0.02K_{T^{\ast }}^{-1}\eta _{10}^{-1}+0.002\ln \eta
_{10}  \label{temfstage1}
\end{equation}%
It is clear that this formula is not applicable if $\eta _{10}<0.35.$ In the
universe with very low baryon density $\left( \eta _{10}\ll 1\right) $ the
neutron concentration never drops below the deuterium concentration. It
freezes-out before. In this case the nucleosynthesis is over very fast after
beginning and neutron concentration freezes-out before the substantial part
of the neutrons is converted into $^{4}He.$ After that the free neutrons
decay. This explains why in the universe with very low baryon density (for
instance, with $\eta _{10}\simeq 10^{-2})$ the helium abundance is less than
one percent (see Fig.3). When I was deriving the formula (\ref{he42}) for $%
^{4}He-$abundance I assumed that the reactions converting the neutrons into $%
^{4}He$ are very efficient and able to transfer most of the available
neutrons into more heavy elements. This means that this formula is valid
only for $\eta _{10}>0.35.$ From the observations of the luminous baryonic
matter we know that $1<\eta _{10}<10^{2}$ and therefore we concentrate from
now on only on this range of parameter $\eta _{10}$. All the derivations
below will be done under this assumption.

The neutron concentration drops to $X_{n}\sim $ $X_{D}\sim R_{1}$ at the
temperature $T^{\ast }$ given by (\ref{temfstage1})$.$ After that the
solution (\ref{sold1}) is not valid anymore and the system quickly gets to
another attractor which correspond to the quasi-equilibrium solution of the
equation (\ref{feq1}), namely, 
\begin{equation}
X_{n}=\frac{1}{R_{1}}X_{D}^{2}\left[ 1+4\frac{X_{n}}{X_{D}}+O\left( \left( 
\frac{X_{n}}{X_{D}}\right) ^{2}\right) \right]   \label{attr}
\end{equation}%
As $X_{n}/X_{D}$ continues to decrease one can neglect the deviations from $%
X_{n}\simeq X_{D}^{2}/R_{1}$ and the equation (\ref{feq2}) for deuterium
takes the following form 
\begin{equation}
\frac{dX_{D}}{dT_{MeV}}=2a\cdot K\left( T\right) \eta _{10}\left(
X_{D}^{2}+2R_{2}X_{D}\right)   \label{d1}
\end{equation}%
Now the deuterium becomes the \textquotedblright key
element\textquotedblright\ and determines its own \textquotedblright
fate\textquotedblright\ regulating simultaneously the quasi-equilibrium
concentrations of the other elements including the neutrons. Since $R_{2}$
practically doesn't change in the relevant temperature interval (see (\ref%
{RR})) it can be treated as a constant and the equation (\ref{d1}) can be
easily integrated: 
\begin{equation}
\left( 1+\frac{2R_{2}}{X_{D}\left( T\right) }\right) =\left( 1+\frac{2R_{2}}{%
X_{D}\left( T^{\ast }\right) }\right) \exp \left( 4a\eta
_{10}R_{2}\int_{T}^{T^{\ast }}K\left( T\right) dT\right) ,  \label{solde}
\end{equation}%
where the temperature is expresses in $MeV.$ When temperature goes to zero $%
\left( T\rightarrow 0\right) $ the deuterium concentration doesn't vanish,
instead it freezes-out at $X_{D}^{f}\equiv X_{D}\left( T\rightarrow 0\right)
.$ Taking into account that $X_{D}\left( T^{\ast }\right) \sim R_{1}\gg R_{2}
$ and estimating the integral in (\ref{solde}) as $\sim K\left( T^{\ast
}\right) T^{\ast },$ where $T^{\ast }$ is given by (\ref{temfstage1}), we
obtain the following expression for the deuterium freeze-out concentration$:$%
\begin{equation}
X_{D}^{f}\simeq \frac{2R_{2}}{\left( \exp \left( A\eta _{10}\right)
-1\right) }  \label{froutd}
\end{equation}%
where 
\begin{equation}
A\equiv 2aR_{2}K\left( T^{\ast }\right) T^{\ast }  \label{coeffA}
\end{equation}%
The numerical coefficient $A$ only slightly varies with $\eta _{10}.$
Actually when $\eta _{10}$ changes by two decade from $1$ to $10^{2}$ this
coefficient increases only twice from $\sim 0.1$ to $\sim 0.2$ . The
expression (\ref{froutd}) fits very well the results of the numerical
calculations presented in Fig.3. If we want to get better accuracy using
analytical approach we can do it taking into account the temperature
dependence of the reaction rates. The formula (\ref{froutd}) is in
satisfactory agreement with the numerical results if we take $A=0.1=const$.
At $\eta _{10}<1/A\sim 10$ the good approximation for (\ref{froutd}) is 
\begin{equation}
X_{D}^{f}\simeq \frac{2R_{2}}{A}\sim 4\times 10^{-4}\eta _{10}^{-1}
\label{appd}
\end{equation}%
We see that in this range of $\eta _{10}$ the deuterium freeze-out
concentration decreases nearly linearly with $\eta _{10}.$ It is easy to
understand. In this case the freeze-out concentration never drops below $%
R_{2}\sim 10^{-5}$ and as it is clear from the equation (\ref{d1}) $DD-$%
reactions always dominate over $Dp-$reactions in destroying deuterium.
Therefore, the deuterium concentration freeze out when $\lambda
_{DD}X_{D}^{f}\sim t^{-1};$ since $\lambda _{DD}\propto n_{B}\propto \eta
_{10}$ we see that in the leading order $X_{D}^{f}$ should be inversely
proportional to $\eta _{10}$ (compare to (\ref{begnucl})).

On the contrary, if $\eta _{10}>10$ the linear in $X_{D}-$term in the
equation (\ref{d1}) dominates after $X_{D}$ drops below $R_{2}\sim 10^{-5}$
and after that $X_{D}\propto \exp \left( -\eta _{10}\times \text{ function
of }T\right) ;$ hence as it follows from (\ref{froutd}) the freeze-out
concentration in this limit is 
\begin{equation}
X_{D}^{f}\simeq 2R_{2}\exp \left( -A\eta _{10}\right)  \label{appd1}
\end{equation}%
and decays by five order of magnitude from $\sim 10^{-5}$ to $\sim 10^{-10}$
when $\eta _{10}$ changes by only one decade from $10$ to $100$ (see Fig.3)$%
. $ In this case the freeze-out concentration is entirely determined by the
reaction $Dp\rightarrow ^{3}He\gamma $ which dominates over $DD-$deuterium
destruction during the last stage before freeze-out. Hence in a dense
universe nearly all deuterium is very efficiently burned down in this
reaction. The deuterium abundance is very sensitive indicator of the baryon
density. This allows us to put rather strong upper bound on $\eta _{10}$
from observations.

\section{Helium-3 and Tritium}

Now we can calculate the freeze-out abundances of the other light elements
using the quasi-equilibrium conditions and assuming that these conditions
are still satisfied at the moment of deuterium freeze-out. First I consider $%
^{3}He$ and assume that $X_{D}^{f}>R_{2}\sim 10^{-5},$ that is, I consider
the case of $1<\eta _{10}<10.$ In this case the deuterium freeze-out is
determined by $DD-$reaction and happens at the time determined by condition $%
\lambda _{DD}X_{D}^{f}\sim t_{D}^{-1}$. The freeze-out time for $^{3}He$ can
be estimated requiring that the reaction $^{3}HeD\rightarrow ^{4}Hen$
becomes inefficient in converting the significant amount of $^{3}He$ into $%
^{4}He.$ This occurs when $\lambda _{^{3}HeD}X_{D}\sim t_{^{3}He}^{-1}.$
Since $\lambda _{^{3}HeD}$ is in few times bigger than $\lambda _{DD}$ it is
clear that $^{3}He$ concentration freezes-out a little bit later than the
deuterium concentration. \ This means that at the moment of deuterium
freeze-out the $^{3}HeD-$reaction is still efficient in returning neutrons
back to \textquotedblright $np-$reservoir\textquotedblright \footnote{%
The same is true for the tritium since $\lambda _{TD}\simeq \lambda
_{He_{3}D}.$ Note also that also $DD-$reaction at this time still continue
actively participate in refilling the \textquotedblright $np-$%
reservoir\textquotedblright\ in accordance with small quasi-equilibrium
neutron demand.}; hence the quasi-equilibrium solution for the neutrons (\ref%
{attr}) derived under this assumption is still valid. Substituting $%
X_{n}=X_{D}^{2}/R_{1}$ in (\ref{qeHe3}) we can express \ the
quasi-equilibrium concentration of $^{3}He$ through $X_{D}:$ 
\begin{equation}
X_{^{3}He}=\frac{3}{2}\frac{\lambda _{DD1}X_{D}+2\lambda _{Dp}X_{p}}{\lambda
_{^{3}HeD}+2\left( \lambda _{^{3}Hen}/R_{1}\right) X_{D}}  \label{He3qeab}
\end{equation}%
After deuterium freezes-out the small leakage from \textquotedblright $D$ to 
$^{3}He-$reservoir\textquotedblright\ is still able to keep stationary
quasi-equilibrium \textquotedblright flow through $^{3}He-$%
reservoir\textquotedblright . Actually in considered case the $^{3}He$%
-concentration is significantly smaller than $X_{D}^{f}$ and the deuterium
demand needed to compensate the leakage of $^{3}He$ through
\textquotedblright $^{3}HeD-$pipe\textquotedblright\ is not very high. After
a short time when the $^{3}HeD-$reaction becomes inefficient both
\textquotedblright $DD1$ and $^{3}HeD-$pipes\textquotedblright\ close up
simultaneously and the $^{3}He-$freeze-out concentration can be obtained
from (\ref{He3qeab}) substituting there $X_{D}^{f}$ given by (\ref{froutd}) 
\begin{equation}
X_{^{3}He}^{f}\sim \frac{0.2X_{D}^{f}+10^{-5}}{1+4\times 10^{3}X_{D}^{f}},
\label{He3frout}
\end{equation}%
Here I first normalized all reaction rates on $\lambda _{^{3}HeD}$ and then
used the experimental values for the obtained ratios. Of course these ratios
depend on the temperature. However, as one can check, they do not change too
much , namely, not more than by factor two in the whole range of the
relevant temperatures for the two decade of baryon density. For the
definiteness I took them at $T\sim 0.06$ $MeV.$ The obtained result is in
excellent agreement with the results of the numerical calculations presented
in Fig.3. We see that if $X_{D}^{f}\simeq 10^{-3}$ then the appropriate $%
^{3}He$ abundance is ten times smaller that the deuterium abundance. It is
clear from the above expressions that this suppression of $^{3}He-$abundance
compared to the deuterium abundance is mostly due to the reactions
converting $^{3}He$ into tritium, which are still very efficient at the
moment of freeze-out because of the rather high concentration of free
neutrons. In more dense universe where the deuterium abundance is smaller
the availability of the free neutrons appropriately reduces. If, for
instance, $X_{D}^{f}$ is smaller than $\sim 2.5\times 10^{-4}$ (see the
denominator in the formula (\ref{He3qeab}) and (\ref{He3frout})) then the
reaction $^{3}Hen\rightarrow Tp$ does not play any significant role in
determining the final $^{3}He-$abundance. This is why we can still use (\ref%
{He3frout}) to estimate $X_{^{3}He}^{f}$even in the universe with a
relatively high baryon density, where $X_{D}^{f}<10^{-5},$ although the free
neutron concentration can significantly deviate from given by (\ref{attr}).

>From (\ref{He3frout}) it follows that, if $X_{D}^{f}\simeq 1.2\times
10^{-5}, $ then the helium-3 freeze-out concentration is equal to the
deuterium concentration. This is in a very good agreement with the numerical
results. In the universe with $\eta _{10}>10$ the helium-3 is produced in
the reaction $Dp\rightarrow ^{3}He\gamma $ and destroyed in the process $%
^{3}HeD\rightarrow ^{4}Hen.$ Irrespective how big is $X_{D}$ these two
competing processes give rise to the final $^{3}He-$abundance $%
X_{^{3}He}^{f}\simeq $ $\lambda _{Dp}/\lambda _{^{3}HeD}\simeq $ $10^{-5}$
even in the case when the deuterium is practically absent. Slight deviation
of the numerical results from predicted here constant $^{3}He-$
concentration in this limit is due to weak temperature dependence of the
reaction rates.

Similar by, one can find that the tritium quasi-equilibrium concentration is
equal to 
\begin{equation}
X_{T}=\left( \frac{3}{2}\frac{\lambda _{DD2}}{\lambda _{DT}}+2\frac{\lambda
_{^{3}Hen}}{\lambda _{DT}}\frac{X_{^{3}He}}{R_{1}}\right) X_{D}  \label{Tf}
\end{equation}%
Using the experimental values for the ratio of the reaction rates $\lambda
_{DD2}/\lambda _{DT}\simeq 0.01$ and $\lambda _{^{3}Hen}/\lambda _{DT}\simeq
1$ we can easily understand the dependence of tritium freeze-out
concentration on $\eta _{10}.$ The formulae (\ref{He3qeab}) and (\ref{Tf})
explain why the $^{3}He-$track in Fig.2$,$ in distinction from $T,$ doesn't
\textquotedblright repeat\textquotedblright\ the track of the deuterium ,
namely, the $^{3}He-$ concentration increases monotonically all the time,
while $T-$concentration first reaches the maximum and then decreases until
it gets its freeze-out value.

\section{Lithium-7 and Beryllium-7\textit{\ }}

The quasi-equilibrium concentrations of $^{7}Li$ and $^{7}Be$ can be
determined from the equations 
\begin{equation}
\frac{1}{12}\lambda _{^{4}HeT}X_{^{4}He}X_{T}+\frac{1}{7}\lambda
_{^{7}Ben}X_{^{7}Be}X_{n}=\frac{1}{7}\lambda _{^{7}Lip}X_{^{7}Li}X_{p}
\label{qeLi7}
\end{equation}%
and 
\begin{equation}
\frac{1}{12}\lambda _{^{4}HeHe_{3}}X_{^{4}He}X_{^{3}He}=\frac{1}{7}\lambda
_{^{7}Ben}X_{^{7}Be}X_{n}  \label{qeBe7}
\end{equation}%
where I took into account only dominating reactions in which, respectively, $%
^{7}Li$ and $^{7}Be$ are produced and destroyed. One can check that if $\eta
_{10}>1$ then the other reactions as, for instance, $^{7}LiD\rightarrow
n+2^{4}He$ and $^{7}BeD\rightarrow p+2^{4}He$ for can be ignored. It
immediately follows from these equations that 
\begin{equation}
X_{^{7}Li}=\frac{7}{12}\frac{X_{^{4}He}}{X_{p}}\left( \frac{\lambda
_{^{4}HeT}}{\lambda _{^{7}Lip}}\right) \left( X_{T}+\frac{\lambda
_{^{4}HeHe_{3}}}{\lambda _{^{4}HeT}}X_{^{3}He}\right)   \label{XLi7}
\end{equation}%
The ratio of $\lambda _{^{4}HeT}/\lambda _{^{7}Lip}$ is remarkably constant
within rather broad temperature interval, namely, it increases from $\sim
2.2\times 10^{-3}$ to $\sim 3\times 10^{-3}$ when the temperature drops in
three times from $0.09$ $MeV$ to $0.03$ $MeV.$ In distinction from that $%
K_{1}\left( T\right) \equiv \lambda _{^{4}He^{3}He}/\lambda _{^{4}HeT}$
varies quite significantly in the same temperature interval: $K_{1}\left(
T\right) \simeq \left( 5\div 0.6\right) \times 10^{-2}$, that, is it drops
nearly ten times when the temperature drops in three times. Substituting $%
X_{^{4}He}\simeq 0.25,$ $X_{p}\simeq 0.75$ in (\ref{XLi7}) we obtain 
\begin{equation}
X_{^{7}Li}\simeq \left( 3\div 5\right) \times 10^{-4}\left(
X_{T}+K_{1}\left( T\right) X_{^{3}He}\right)   \label{XLi7a}
\end{equation}%
To get the freeze-out concentration of $^{7}Li$ we have to substitute in
this formula the appropriate value of $X_{T}$ , $K_{1}\left( T^{\ast
}\right) $ and $X_{^{3}He}$ at the moment when $^{7}Li-$freeze-out. This
moment can be evaluated analyzing the freeze-out condition for $^{7}Li,$ $%
^{7}Be-$reactions. If $X_{D}>3\times 10^{-5}$ ($1<\eta _{10}<5$) this
freeze-out occurs after the deuterium gets its final abundance. Therefore to
estimate $X_{^{7}Li}^{f}$ in this case one can substitute the obtained above
values for $X_{^{3}He}^{f}$ and $X_{T}^{f}$ in (\ref{XLi7a}). If $\eta
_{10}=1$ the first term there dominates. Taking into account that $%
X_{T}^{f}\sim 0.01\times X_{D}^{f}\simeq 4\times 10^{-6}$ we get $%
X_{^{7}Li}^{f}\left( \eta _{10}=1\right) \simeq 10^{-9}.$ As $\eta _{10}$
increases $X_{T}^{f}$ drops and therefore $X_{^{7}Li}^{f}$ also decreases
until the second term in (\ref{XLi7a}) starts to dominate. After that the $%
^{7}Li$ abundance starts to increase. This increase is due to two reasons.
First of all it comes from the freeze-out temperature dependence of $K\left(
T^{\ast }\right) ,$ which can be easily understood. Namely, the freeze-out
temperature for $^{7}Li$ in this case is mostly determined by the efficiency
of $^{7}Ben-$reaction which in turns depends on the neutron concentration.
In more dense universe the deuterium and free neutrons burned down more
efficiently and disappear earlier (at higher temperature) than in the
universe with small baryon density. Therefore the $^{7}Li$ concentration
freezes-out at higher temperatures for which $K_{1}$ is bigger. Second
reason is the following. If $\eta _{10}>5$ the $^{7}Ben-$reaction become
inefficient before $^{3}He$ reaches its final freeze-out concentration.
Therefore, to estimate $X_{^{7}Li}^{f}$ in this case we have to substitute
in (\ref{XLi7a}) the actual value of $X_{^{3}He}$ at the moment when the $%
^{7}Li-$concentration freezes-out, which is bigger than $X_{^{3}He}^{f}.$
The numerical calculation show that after passing through a relatively deep
minimum $X_{^{7}Li}^{f}\left( \eta _{10}=2\div 3\right) \simeq 10^{-10}$ the
Lithium concentration comes back to $\sim 10^{-9}$ at $\eta _{10}\simeq 10.$
The \textquotedblright trough\textquotedblright\ in $X_{^{7}Li}^{f}-\eta
_{10}$ dependence has a very simple explanation, namely, it is due to the
competition of two reactions. In the universe with $\eta _{10}<2\div 3$ most
of the Lithium-7 is produced directly as a result of $^{4}HeT-$interactions.
The efficiency of this process decreases with increase of $\eta _{10}$ and
as $\eta _{10}$ becomes bigger than about $2\div 3$ the reaction $^{7}Ben$
takes over compared to the direct $^{7}Li-$production. In this case most of
the Lithium-7 is produced via intermediate \textquotedblright $^{7}Be-$%
reservoir\textquotedblright 

The Beryllium-7 is not so important from the observational point of view.
Therefore, to get an idea about its expected concentration we estimate the
amount of $^{7}Be$ only for $1<\eta _{10}<5.$ In this case $^{7}Be$
freezes-out after deuterium and the quasi-equilibrium solution (\ref{attr})
for the free neutrons is still valid at this time. Hence we get 
\begin{equation}
X_{^{7}Be}^{f}=\frac{7}{12}\frac{X_{^{4}He}}{\left( X_{D}^{f}\right) ^{2}}%
R_{1}\left( \frac{\lambda _{^{3}He^{4}He}}{\lambda _{^{7}Ben}}\right)
X_{^{3}He}^{f}\sim O\left( 1\right) 10^{-12}\frac{X_{^{3}He}^{f}}{\left(
X_{D}^{f}\right) ^{2}}  \label{be7f}
\end{equation}%
where I used the experimental values for the ratios of the appropriate
reactions; here the product of these ratios changes in about five times in
the relevant temperature interval. $\ $At $\eta _{10}=1$ we have $%
X_{D}^{f}\sim 4\times 10^{-4}$, $X_{^{3}He}^{f}\sim 0.1X_{D}^{f}$ and
correspondingly $X_{^{7}Be}^{f}\sim 2.5\times 10^{-10}.$

\section{Conclusions}

The derived above abundances of the light elements are in very good
agreement with the results of the numerical calculations reviewed, for
instance, in \cite{reviews},\cite{OSW}. The analytical derivation is useful
because it allow us to look inside \textquotedblleft black box" (computer)
and understand an interesting physics (for instance, attractors behavior
etc.) taking place during nucleosynthesis. Without referring to the computer
codes one can estimate the final abundances of the light elements and
understand their dependences on the main cosmological parameters. The
theoretical calculations of the abundances are in agreement with
observations \cite{reviews},\cite{OSW} and the estimates of the baryon
density in the universe from nucleosynthesis come in impressive
correspondence with CMB temperature fluctuation measurements. This gives
strong support to the standard cosmological model.

\end{document}